

\magnification=1200\baselineskip=0.33truein

\vskip1truein

\centerline{SUPPRESSION OF BREMSSTRAHLUNG AT NON-ZERO TEMPERATURE}

\bigskip
\centerline{H. Arthur Weldon}

\centerline{Department of Physics}

\centerline{West Virginia University}

\centerline{Morgantown, WV 26506-6315}

\centerline{July 26, 1993}

\bigskip

The  first-order bremsstrahlung emission spectrum  is $\alpha d\omega/\omega$
at zero temperature. If the radiation is emitted into a region that  contains a
thermal
 distribution of photons, then the rate is increased by a factor $1+N(\omega)$
where
$N(\omega)$ is the Bose-Einstein function. The stimulated emission changes the
spectrum  to $\alpha
Td\omega/\omega^{2}$ for $\omega\ll T$. If this were correct, an infinite
amount of energy would be
radiated in the low frequency modes. This unphysical result indicates a
breakdown of perturbation
theory. The paper computes the bremsstrahlung rate to all orders of
perturbation theory, neglecting
the recoil of the charged particle. When the perturbation series is summed, it
has a different
low-energy behavior. For $\omega\ll\alpha T$, the spectrum is   independent of
$\omega$ and has a
value proportional to $d\omega/\alpha T$ .

\vfill\eject

\bigskip

\centerline{I. INTRODUCTION AND SUMMARY}

\centerline{A. Background}

In studies of the quark-gluon plasma to be produced in ultrarelativistic heavy
ion collisions and,
more generally, in studies of field theory at finite temperature, a central
concern is
how the cancellation of infrared divergences  affects various finite, physical
quantities.
The experimentally measured rates for certain low energy processes can be
significantly
modified by the infrared structure.

This paper will investigate the emission of low energy photons  by a charged
particle that
passes through a fixed-temperature plasma. The charged particle must undergo a
collision in order
to radiate. If the radiated energy is much smaller than the energy transfer in
the collision then the
inelastic cross section factors:
$$2\omega{d\sigma\over d^{3}kdq^{2}}\approx 2\omega{dP(q^{2})\over d^{3}k}
{d\sigma\over dq^{2}}
\hskip0,5truein (|k\cdot q|\ll |q^{2}|)\eqno(1.1)$$
with the collision cross section $d\sigma/dq^{2}$ independent of $k\cdot q$,
where $q=p-p^{\prime}$.
Thus to investigate the probability of radiation, $P$, the details of the hard
scattering do not
matter. All that matters is that the charged particle began with four-momentum
$p^{\mu}$ and ended
with four-momentum $p^{\prime\mu}$.
The radiation probability to first order in $\alpha$ is
$$2\omega{dP_{1}\over d^{3}k}={1\over (2\pi)^{3}}\sum_{\rm
pol}|\epsilon_{\mu}\cdot J^{\mu}|^{2}
\ [1+N(\omega)]\eqno(1.2)$$
where $\omega=|\vec{k}|$ and $N(\omega)=[1-\exp(\omega/T)]^{-1}$ is the
Bose-Einstein function.
When the radiated enegy is small, the matrix element of the electromagnetic
current  is
$$J^{\mu}(k)=ie\left({p^{\prime\mu}\over p^{\prime}\cdot k}-{p^{\mu}\over
p\cdot k}\right)
\ e^{-k/2\Lambda}\eqno(1.3)$$
where
$\Lambda$ is a momentum-space regulator that will be necessary later.
The current results from an on-shell charge of four-momentum $p$ radiating an
on-shell photon of
four-momentum $k$, which gives a Feynman denominator  $(p-k)^{2}-m^{2}=-2p\cdot
k$.
For low-energy radiation this current is valid regardless of the spin of the
charged particle
[1-3]. Because of the denominators in (1.3) the radiation is mostly parallel to
$\vec{p}$ or
$\vec{p}^{\ \prime}$.
When (1.2) is integrated over photon angles, the probability of radiating
energy $\omega$ in any
direction is $${dP_{1}\over d\omega}={A\over\omega} [1+N(\omega)]\
e^{-\omega/2\Lambda}\eqno(1.4)$$
The charged particle momenta only occur in the function $A$:
$$A(p\cdot p^{\prime})
= {\alpha\over \pi}\left({1\over v}\ln({1+v\over 1-v})-2\right)\eqno(1.5)$$
where $v$ is defined by $p\cdot p^{\prime}=m^{2}(1-v^{2})^{-1/2}$.
In terms of momentum transfer $Q$, the limiting behaviors of $A$ are [1,2]
$$A\to{2\alpha\over 3\pi}\ {Q^{2}\over m^{2}}\hskip0.5truein  {\rm for}\ Q\ll
m\eqno(1.6a)$$
$$A\to {2\alpha\over \pi}\ \ln({Q^{2}\over m^{2}}) \hskip0.3truein {\rm for}\
Q\gg m\eqno(1.6b)$$

Formula (1.4) is  classical, except for  photon quantum statistics which
produce
the factor $N$.  The formula fails at low energies because it predicts that the
total
energy radiated will be infinite:
$$\int_{0}^{E_{\rm max}}d\omega {dP_{1}\over d\omega}= \infty\eqno(1.7)$$
regardless of the value of $E_{\rm max}$.
This does not occur at zero temperature, where  the radiation probability is
$A/\omega$; but at
non-zero temperature the  Bose-Einstein factor  makes it more singular:
$${dP_{1}\over d\omega}\approx {AT\over \omega^{2}}\hskip.5truein (\omega\ll
T)\eqno(1.8)$$
The breakdown comes from the smallest values of $\omega$, where the classical
current
approximation (1.3) works best.
The resolution of the breakdown must  occur from a higher order calculation.

\vfill\eject
\centerline{B. Higher Orders}

 To go to higher orders in $\alpha$ one must quantize the electromagnetic field
with the interaction
 $${\cal L}_{I}=-J^{\mu}(x)A_{\mu}(x)\eqno(1.9)$$
with $J^{\mu}$ the classical current (1.3) and $A_{\mu}$ the quantized field.
This interaction will produce multiple emissions and absorptions of real
photons as well as closed
loops of virtual photons.
For example, to second order in $\alpha$
 there are three types of contributions to the  probability  of radiating
energy
$\omega$: either radiate two photons whose energy totals $\omega$; or radiate
one photon of energy
$\omega + \omega^{\prime}$ and absorb another whose energy  is
$\omega^{\prime}$, giving a net enegy
of $\omega$; or radiate a single photon of
energy $\omega$ with a one-loop correction. The infrared divergences
cancel among these processes and give a finite answer $dP_{2}/d\omega$ to order
$\alpha^{2}$.

The semiclassical interaction (1.9) does not conserve energy-momentum: it
allows the particle of
momentum $p^{\mu}$ to radiate and still have momentum $p^{\mu}$ (provided the
hard collision
eventually deflects the charge to $p^{\prime\mu}$) . This is a sensible
approximation if each photon
energy (real or virtual) is much smaller than the energy of the charged
particle.
 To maintain this consistently in
higher orders it is convenient to introduce  the momentum cutoff  $\Lambda$ in
(1.3)
and require $\Lambda\ll(\vec{p}^{2}+m^{2})^{1/2}$ and
$\Lambda\ll(\vec{p}^{\prime 2}+m^{2})^{1/2}$.

Because the current is classical, one can compute the generating funtional for
the
multi-photon Green functions by performing a Gaussian functional integral. This
allows one to
calculate the exact multi-photon amplitudes ${\cal M}$.
The probability of radiatiating a \underbar{net} energy $\omega$
to all orders in perturbation theory is
$$\eqalign{{dP\over d\omega}=\sum_{n=1}^{\infty}\int d\Phi_{1}...d\Phi_{n}
&\delta(k_{1}^{0}+...k_{n}^{0}-\omega)\cr
&{1\over n!}\sum_{\rm pol}|{\cal M}(k_{1},...k_{n})|^{2}\cr}\eqno(1.10)$$
Positive $k^{0}$'s correspond to emission of photons; negative $k^{0}$'s
correspond to absorption
of photons. The phase space integration includes the appropriate statistical
factors:
$$d\Phi={d^{3}k\over 2k(2\pi)^{3}}\times\cases{1+N(k) & $k^{0}=+k$\cr
N(k) & $k^{0}=-k$\cr}\eqno(1.11)$$
The amplitudes ${\cal M}$ contain infrared divergences from the virtual photon
integrations, but the integration over the real photons in (1.10) produces an
infrared-finite
probabilty. This is the Bloch-Nordsieck cancellation at $T\neq 0$. The
cancellation
is very delicate: It  would not occur if negative
$k^{0}$'s (i.e. energy absorption from the heat bath) were omitted from the
delta function in (1.10).

\bigskip
\centerline{C. Summary of Results}

Remarkably, one can calculate the probability (1.10) with only the
approximation $T\ll \Lambda$.
The final answer obtained at the end of Sec. III is
$${dP\over d\omega}=|\Gamma({A\over 2}+i{\omega\over 2\pi T})|^{2}\
{e^{\omega/2T}e^{-|\omega|/\Lambda}\over 4\pi^{2}T\ \Gamma(A)}\ \left({2\pi
T\over
\Lambda}\right)^{A}\eqno(1.12)$$
Here $A\propto \alpha$ is the same function of $p\cdot p^{\prime}$ as (1.5).
This result also applies if $\omega$ is negative, which means that the total
emitted energy
is less than the total absorbed energy.
For $\omega=-\widetilde{\omega}<0$, it has the property
 $${dP\over d\omega}=e^{-\widetilde{\omega}/T}\ {dP\over
d\widetilde{\omega}}\eqno(1.13)$$
In the first order result (1.4), this corresponds to
$\exp(-\widetilde{\omega}/T)
[1+N(\widetilde{\omega})]=N(\widetilde{\omega})$.
The zero temperature limit of (1.12) is discussed in Appendix A.

At $\omega=0$ the exact result (1.12) is finite. This resolves the total energy
problem that arose
in (1.7). The most interesting feature of (1.12)  is that
it involves two dimensionless quantities: $A$ and $\omega/T$.
To display a simpler form  it is helpful to
assume $A\ll 1$.  This is quite reasonable because extremely large momentum
transfers are
necessary to make $A$ large (for example,  $A=1/16$ requires  $Q/m=10^{3}$ ).
Even with $A\ll 1$ the behavior of (1.12) depends on the size of $\omega/T$.
If $A\pi\ll \omega/T$ then
$${dP\over d\omega}\approx{A\over\omega} [1+N(\omega)]\ e^{-\omega/\Lambda}
\hskip0.3truein (A\pi T\ll\omega)\eqno(1.14)$$
which agrees with  (1.4). However this does not hold at the smallest values of
$\omega$.
To approximate (1.12) for $A\ll 1$ and $\omega\ll T$
 one can use
$\Gamma(z)\approx 1/z$ when $|z|\ll 1$. Then
$${dP\over d\omega}\approx{AT\over \omega^{2}+(A\pi T)^{2}}\hskip0.5truein
(\omega\ll T)\eqno(1.15)$$
Naturally (1.15) agrees with  (1.14) in the region of overlap,
$A\pi T\ll\omega\ll T$. But for $\omega\ll A\pi T$ the radiation is suppressed
relative to
the first order rate.  Fig. 1  compares (1.12) with the first-order result over
a wide range of
$\omega$.

Naturally the net radiated energy is finite,
$$\int_{0}^{T}d\omega\ \omega{dP\over d\omega}\approx AT\tan^{-1}({1\over
A\pi})\eqno(1.16)$$
and numerically small.
In the denominator of (1.15)
 the quantity $A\pi T$ appears to be some type of thermal
mass of order $e^{2}T$. However, the folowing discussion will argue that $A\pi
T=\Gamma/2$, where
$\Gamma$ is
 the radiation-damping rate.

It is necessary to emphasize several points: (i) At $T=0$ the exact
bremsstrahlung probability
 does not enjoy this damping. It is peculiar to $T\ne 0$. (ii)  Whenever $T>0$
the suppression will
occur for a charge of any energy.  Whether the charge is relativistic or
nonrelativistic alters the
value of $A$, but the plateau shown in Fig. 1 will always exist.

\bigskip\centerline{D. Interpretation of the Suppression}

One can understand the suppression of low frequency bremsstrahlung as the
radiation reaction
brought about by  unitarity. First, we examine the failure of the  first-order
bremmstrahlung
probability by writing it  as
$$\eqalign{{dP_{1}\over d\omega}&={|F_{1}(\omega)|^{2}\over 2\pi}\cr
F_{1}(\omega)&=\sqrt{2\pi A}\left({1+N(\omega)\over\omega}\right)^{1/2} \
e^{-|\omega|/2\Lambda}\cr}
\eqno(1.17)$$
Let the Fourier transform of this be
$$\tilde{F}_{1}(t)=\int_{-\infty}^{\infty}{d\omega\over2\pi}\ e^{-i\omega t}\
F_{1}(\omega)\eqno(1.18)$$ Then the bremsstrahlung probability has a
time-dependence
$${dP_{1}\over dt}=|\tilde{F}_{1}(t)|^{2}\eqno(1.19)$$
As $t\to \infty$ the Fourier transform of (1.17) behaves like
$$\tilde{F}_{1}(t)\to -i\sqrt{\Gamma}\eqno(1.20)$$
where $\Gamma\equiv 2\pi AT$.
Therefore
$dP_{1}/dt\to \Gamma$ as $t\to\infty$. It is obviously unphysical for the
radiation probability to
remain constant after an infinitely long time.
This disease is familiar in elementary atomic physics
calculations of radiative transitions (e.g. $2p\to 1s+\gamma$ in hydrogen). The
resolution is that at
higher orders the transition amplitude usually falls exponentially with time.
Hence a reasonable
guess is that higher order calculations should replace (1.20) by
 $$\tilde{F}(t)\buildrel ?\over =
-i\theta(t)\sqrt{\Gamma}\ e^{-\Gamma t/2}\eqno(1.21)$$ The Fourier transform to
frequency gives
$$F(\omega) \buildrel ?\over= {\sqrt{\Gamma}\over \omega
+i\Gamma/2}\eqno(1.22)$$
The guess gives a radiation probability
$${dP\over d\omega}\buildrel ?\over = {1\over 2\pi}\ {\Gamma\over
\omega^{2}+(\Gamma/2)^{2}}\eqno(1.23)$$
which coincides with (1.15) the weak-coupling, low-energy limit. Note that this
radiation damping
does not come from modifications to the charged particle trajectory.

The remainder of the paper describes the calculation that yields (1.12). Sec.
II formulates the
bremsstrahlung problem in detail and Sec. III performs the necessary
integrations.
Appendix A discusses the $T=0$ limit of (1.12) and its relation to conventional
$T=0$ calculations.
Appendix B discusses the relation of (1.12) to some previous work of mine.

\bigskip

\bigskip\centerline{II. THERMAL BREMSSTRAHLUNG}

When charged particles are described by classical currents $J^{\mu}$,
quantizing the
radiation field becomes elementary. The generating functional for multi-photon
Green functions is
$$Z(J)=\int D[A]\exp[i\int_{C} d^{4}x({\cal L}_{0}-J\cdot A)]\eqno(2.1)$$
where $C$ is a contour in the complex $x^{0}$ plane that incorporates the
temperature.
For a real-time formulation [4-6] this gives
$$Z(J)=\exp[{-i\over 2}\int {d^{4}k\over
(2\pi)^{4}}J^{\mu}(k)D_{\mu\nu}(k)J^{\nu}(-k)]\eqno(2.2)$$
where the finite-temperature propagator is
$$D_{\mu\nu}(k)=-g_{\mu\nu}
\left({1\over k^{2}+i\epsilon}-i2\pi \delta(k^{2})N\right)\eqno(2.3)$$
with $N=[\exp(|\vec{k}|/T)-1]^{-1}$.
The amplitude for  one real photon is
$$\eqalign{{\cal M}(k)&=-ik^{2}\epsilon_{\mu}(k)
{\delta Z(J)\over \delta J^{\mu}(-k)}\ (2\pi)^{4}\cr
&=\epsilon_{\mu}(k)J^{\mu}(k)\ Z(J)\cr}\eqno(2.4)$$
This amplitude describes emission if $k^{0}>0$; absorption, if $k^{0}<0$.
The amplitude for $n$ photons is
$${\cal M}(k_{1},k_{2},...k_{n})=Z(J)\
\prod_{\ell=1}^{n}[\epsilon_{\mu}(k_{\ell})J^{\mu}(k_{\ell})]\eqno(2.5) $$
To compute probabilities, one must square the amplitude and integrate over
photon momenta.
For the photon phase space it is convenient to employ the notation
$$d\Phi_{\ell}={d^{4}k_{\ell}\over
(2\pi)^{3}}\delta(k_{\ell}^{2})\
[\theta(k_{\ell}^{0})+N(|\vec{k}_{\ell}|)]\eqno(2.6)$$
This correctly weights photon emission  with the statistical factor $1+N$; and
photon absorption  with the factor $N$. The probability of radiating a net
energy $\omega$  is
$$\eqalign{{dP\over d\omega}=\sum_{n=1}^{\infty}\int d\Phi_{1}...d\Phi_{n}
&\delta(k_{1}^{0}+...k_{n}^{0}-\omega)\cr
&{1\over n!}\sum_{\rm pol}|{\cal M}(k_{1},...k_{n})|^{2}\cr}\eqno(2.7)$$
The polarization sums give
$$\sum_{\rm pol}|{\cal M}(k_{1},...k_{n})|^{2}=|Z(J)|^{2}
\prod_{\ell=1}^{n}J_{\mu}(k_{\ell})J^{\mu}(k_{\ell})\eqno(2.8)$$
It is important to note that the current (1.3) has the properties
$$\eqalign{J_{\mu}(k)J^{\mu *}(k)&=J_{\mu}(k)J^{\mu}(-k)\cr
&=-J_{\mu}(k)J^{\mu}(k)<0\cr}\eqno(2.9)$$

{\sl Virtual Photons:} The integration over all  virtual photons is contained
in
the multiplicative factor $|Z(J)|^{2}$. Using (2.2) and (2.3)  gives
$$|Z(J)|^{2}=\exp(V)$$
$$V= -\int {d^{3}k\over 2k(2\pi)^{3}}J_{\mu}(k)J^{\mu}(k)[1+2N]\eqno(2.10)$$
Because $J\sim 1/k$, this integral contains both linear and logarithmic
divergences in the infrared.

{\sl Real Photons:} The delta function that constrains the real photon energies
 can be
represented as $$\delta(k_{1}^{0}+...k_{n}^{0}-\omega)
=\int_{-\infty}^{\infty}{dz\over 2\pi}e^{-i\omega z}\
e^{i(k_{1}^{0}+...k_{n}^{0})z}\eqno(2.11)$$
In (2.7) each integration over $d^{4}k_{\ell}$ will give the same function of
$z$:
$$R(z)=\int d\Phi e^{ik^{0}z}\sum_{\rm
pol}|\epsilon_{\mu}(k)J^{\mu}(k)|^{2}\eqno(2.12)$$
More explicity this is
$$\eqalign{R(z)=\int{d^{3}k\over
2k(2\pi)^{3}}&J_{\mu}(k)J^{\mu}(k)\cr
&\times\left([1+N]e^{ikz}+Ne^{-ikz}\right)\cr}\eqno(2.13)$$ This is the
contribution of the real photons. Because $J\sim 1/k$,  this integral has both
linear and logarithmic
infrared divergence.

The probability (2.7) is $$\eqalign{{dP\over d\omega}&=
|Z(J)|^{2}\int_{-\infty}^{\infty}{dz\over 2\pi}e^{-i\omega z}
\sum_{n=1}^{\infty}{[R(z)]^{n}\over
n!}\cr &=\int_{-\infty}^{\infty}{dz\over 2\pi}e^{-i\omega z}
\exp[\overline{R}(z)]\cr}\eqno(2.14)$$
(the $n=0$ term has no Fourier transform at $\omega\neq 0$) and $\overline{R}$
is
$$\eqalign{\overline{R}(z)&=R(z)+V\cr
&=\int{d^{3}k\over 2k(2\pi)^{3}}J_{\mu}(k)J^{\mu}(k)\cr
&\hskip0.2truein\times\left([1+N]e^{ikz}
+Ne^{-ikz}-[1+2N]\right)\cr}\eqno(2.15)$$
Each term in $\overline{R}$ has a clear interpretaion. Recall that $z$ is the
variable conjugate to
the net energy $\omega$. In (2.15) the term proportional to $1+N$ represents
the stimulated
emission of energy; the term proportional to $N$ represents the absorption of
energy;
the subtracted term $1+2N$ represents all virtual photons (emitted and
absorbed).
At small $k$, $J^{\mu}\sim 1/k$ and $N\sim 1/k$.
Expanding out the $\exp(\pm ikz)$ for small $k$ shows that   the linearly
divergent terms
$dk/k^{2}$ cancel and the logarithmically divergent terms $dk/k$ cancel. Thus
$\overline{R}$ is a completely finite function. (See (3.15) for
$\overline{R}$.)

\bigskip

\bigskip\centerline{III. EXPLICIT INTEGRATION}

\centerline{A. Angular Integration}

The remaining task is to perform the integrations necessary for the probability
(2.14).
The first step is to write
$$\eqalign{\overline{R}(z)=&\int_{0}^{\infty}{dk\over k}A\cr
&\times\left([1+N]e^{ikz}
+e^{-ikz}-[1+2N]\right)\cr}\eqno(3.1)$$
where $A$ contains the angular integration:
$$A={k^{2}\over 2}\int {d\Omega\over(2\pi)^{3}}J_{\mu}(k)J^{\mu}(k)\eqno(3.2)$$
Using $k^{\mu}=k(1,\hat{k})$ in  the  current (1.3) gives
$$\eqalign{J_{\mu}(k)J^{\mu}(k)&={e^{2}\over k^{2}}
\Bigl[{2(E^{\prime}E-\vec{p}^{\ \prime}\cdot
\vec{p})\over(E^{\prime}-\vec{p}^{\ \prime}\cdot
\hat{k})(E-\vec{p}\cdot \hat{k})} \cr
& -{m^{2}\over (E^{\prime}-\vec{p}^{\ \prime}\cdot \hat{k})^{2}}
-{m^{2}\over (E-\vec{p}\cdot \hat{k})^{2}}\Bigr]
e^{-k/\Lambda}\cr}\eqno(3.3)$$
 The factors of $k$ completely cancel  so that $A$
is independent of $k$. This is the same angular integration  that arises in
lowest order
bremsstrahlung and is described in textbooks [2]. It turns out that $A$ is a
rather messy function of
momentum transfer $Q$. It is a bit simpler to express it in terms of
$\sigma=p^{\prime}\cdot
p/m^{2}$:
$$A={2\alpha\over\pi}\left({\sigma\over\sqrt{\sigma^{2}-1}}\ln\Bigl(\sigma+\sqrt{\sigma^{2}-1}
\Bigr)-1\right)\eqno(3.4)$$
As noted by Weinberg [7], it is simplest to express $A$ in terms of the
relative velocity $v$ of the final particle in the rest frame of the initial
particle (or vice
versa) defined by $p^{\prime}\cdot p=m^{2}(1-v^{2})^{-1/2}$ so that
 $$A(p\cdot p^{\prime})
= {\alpha\over \pi}\left({1\over v}\ln({1+v\over 1-v})-2\right)\eqno(3.5)$$

\bigskip

\centerline{B.  Integration Over $k$}

To perform the $k$ integration necessary for (3.1),
 separate $\overline{R}$  into a temperature-independent part and a
temperature-dependent
part:
$$\overline{R}(z)=\overline{R}_{0}(z)+\overline{R}_{T}(z)\eqno(3.6a)$$
$$\overline{R}_{0}(z)=A\int_{0}^{\infty}{dk\over k}
(e^{ikz}-1)\exp(-k/\Lambda)\eqno(3.6b)$$
$$\overline{R}_{T}(z)=2A\int_{0}^{\infty}{dk\over k}
{\cos(kz)-1\over\exp(k/T)-1}\exp(-k/\Lambda)\eqno(3.6c)$$
Both $\overline{R}_{0}$ and $\overline{R}_{T}$ are infrared finite.
For $\overline{R}_{0}(z)$, expand the integrand in powers of $z$
$$\overline{R}_{0}(z)=A\sum_{n=1}^{\infty}{(iz)^{n}\over n!}\int_{0}^{\infty}dk
k^{n-1}\
e^{-k/\Lambda}\eqno(3.7)$$ The displayed integrals over $k$ each give
$(n-1)!\Lambda^{n}$. The sum on
$n$ is  elementary:
 $$\overline{R}_{0}(y)=-A\ln[1-i\Lambda z]\eqno(3.8)$$
For $\overline{R}_{T}(z)$, put $k=2Tx$ so that
$$\overline{R}_{T}(z)=A\int_{0}^{\infty}{dx\over x}{\cos(2Tzx)-1\over
\sinh(x)}e^{-x[1+2\ T/\Lambda]}
\eqno(3.9)$$
This is finite at $x=0$ but it is convenient to multiply the integrand by an
additional
convergence factor $x^{\mu}$ in order to use [8]
$$\int_{0}^{\infty}dx x^{\mu-1}{e^{-\beta x}\over \sinh(x)}
=2^{1-\mu}\Gamma(\mu)\ \zeta[\mu,{\beta+1\over 2}]\eqno(3.10)$$
This gives
$$\overline{R}_{T}(z)=\lim_{\mu\to 0}A2^{1-\mu}\Gamma(\mu)\ {\rm Re}\
\Bigl(\zeta[\mu,\hat{q}]-\zeta[\mu,q]\Bigr) \eqno(3.11)$$
where $\hat{q}=q+iTz$ and $q=1+T/\Lambda$.
Now expand  $\zeta$ in a Taylor series about $\mu=0$ using [8]
$$\zeta[\mu,q]|_{\mu=0}=-q+{1/2}\eqno(3.12a)$$
$${d\zeta[\mu,q]\over d\mu}|_{\mu=0}=\ln[\Gamma(q)]-{1\over
2}\ln(2\pi)\eqno(3.12b)$$
When the limit $\mu\to 0$ is taken in (3.11) the result is
$$\overline{R}_{T}(z)=A\ln\left({\Gamma(q+iTz)\Gamma(q-iTz)\over\Gamma(q)\Gamma(q)}\right)\eqno(3.13)$$
$\overline{R}(z)$ is the sum of (3.8) and (3.13).

\bigskip
\centerline{C.  Integration Over $z$}

The  last integral to perform   is the Fourier transform
$${dP\over d\omega}=\int_{-\infty}^{\infty}{dz\over 2\pi}\ e^{-i\omega z}
\exp(\overline{R}(z))\eqno(3.14)$$
where the integrand is
$$\exp(\overline{R}(z))= \Bigl[{\Gamma(q+iTz)\Gamma(q-iTz)
\over (1-i\Lambda z)\Gamma(q)\Gamma(q)}\Bigr]^{A}\eqno(3.15)$$
 and $q=1+T/\Lambda$. Since $A$ is non-integer, (3.15) has an infinite number
of branch cuts in the
variable $z$. For $\omega>0$ (corresponding to net emission of energy
by the charge) one can close the $z$ contour in the lower half plane. Then
there
is a branch point at $z_{0}=-i/\Lambda$  and
an infinte set of  branch points   at
$z_{n}=-i/\Lambda-in/T$ for n=1,2,3... that come from the poles of
$\Gamma(q-iTz)$.
Choose the branch cuts to run from $z_{n}$ to $z_{n}+\infty$ as shown in Fig. 2
so that
$${dP\over d\omega}=\sum_{n=0}^{\infty}\int_{z_{n}}^{z_{n}+\infty}{dz\over
2\pi}\ e^{-i\omega z}
{\rm Disc}_{n}\Bigl[\exp(\overline{R}(z))\Bigr]\eqno(3.16)$$
The discontinuity across the n'th branch cut is
$${\rm Disc}_{n}\Bigl[\exp(\overline{R}(z))\Bigr]=[1-e^{-i2\pi
A}]\exp(\overline{R}(z))\eqno(3.17)$$
Then put $z=z_{n}+r$ where $r$ is real:
$$\eqalign{{dP\over d\omega}=[1-&e^{-i2\pi A}]\sum_{n=0}^{\infty}
e^{-i\omega z_{n}}\cr
&\int_{0}^{\infty}{dr\over 2\pi}\ e^{-i\omega
r}\exp(\overline{R}(z_{n}+r))\cr}\eqno(3.18)$$
The value of the function along the branch cut is
$$\exp(\overline{R}(z_{n}+r))= \Bigl[C_{n}(r){\pi T\
e^{i\pi(-n+1/2)}\over \Lambda \sinh(\pi Tr)}\Bigr]^{A}\eqno(3.19a)$$
$$C_{n}(r)={\Gamma(1+2T/\Lambda+n+iTr)\over
\Gamma(1+n+iTr)[\Gamma(1+T/\Lambda)]^{2}}
\eqno(3.19b)$$
 At this stage it is necessary to make the
approximation $T\ll \Lambda$ so that  $C_{n}(r)\to 1$.
This allows  the summation on $n$ in (3.18) to be done:
$${dP\over d\omega}={e^{\omega/2T}\ e^{-\omega/\Lambda}\sin(A\pi)\over\sin(
A\pi/ 2-i\omega
/2T)}\left({\pi T\over \Lambda}\right)^{A}\ {I(\omega)\over 2\pi}\eqno(3.20)$$
$$I(\omega)=\int_{0}^{\infty}dr e^{-i\omega r}[\sinh(\pi Tr)]^{-A}\eqno(3.21)$$
This is a known integral [8]
$$I(\omega)={\Gamma(1-A)\Gamma(A/2+i\omega/2\pi T)
\over \pi T 2^{1-A}\Gamma(1-A/2+i\omega/ 2\pi T)}\eqno(3.22)$$
Using the reflection property of the gamma function gives the final result
quoted in (1.12):
$${dP\over d\omega}={e^{\omega/2T}\ e^{-\omega/\Lambda}\over
\Gamma(A)4\pi^{2}T}
\left({2\pi T\over \Lambda}\right)^{A}|\Gamma({A\over 2}+i{\omega\over 2\pi
T})|^{2}\eqno(3.23)$$

\bigskip
\centerline{ACKNOWLEDGMENT}

This work was supported in part by  the U.S. National Science Foundation under
grant PHY-9213734.

\bigskip
\centerline{APPENDIX: THE ZERO-TEMPERATURE LIMIT}

 At zero temperature (1.12) has the simple behavior
 $${dP\over d\omega}= {1\over\omega}\left({\omega\over \Lambda}\right)^{A}\
{e^{-\omega/\Lambda}\over \Gamma(A)}\eqno(A1)$$
and easily satisfies
$$\int_{0}^{\infty}d\omega{dP\over d\omega}=1\eqno(A2)$$
Since all  detectors have some energy threshold
for detection, physical probabilities must  include an integration over the
below-threshold
photons. If the detector measures total radiant energy deposition (as in a
calorimeter)
above $E_{\rm min}$  then
  the  appropriate probability is
$$\sum_{n=1}^{\infty}\int\prod_{\ell=1}^{n}\Bigl[{d^{3}k_{\ell}\over 2k_{\ell}
(2\pi)^{3}}\Bigr]
\ \theta(E_{\rm min}-\sum_{j}k_{j})
{|{\cal M}_{n}|^{2}\over n!}\eqno(A3)$$
This probability is
$$\int_{0}^{E_{\rm min}}d\omega {dP\over d\omega}=\left({E_{\rm min}\over
\Lambda}\right)^{A}
{1\over \Gamma(1+A)}\eqno(A4)$$
with $\exp(-\omega/\Lambda)$ neglected.

The more familiar situation is a detector that is sensitive to single photons
each of which has energy
above some threshold $E_{\rm min}$. In that case the single theta function in
(A3) is replaced by a
product: $$\theta(E_{\rm min}-k_{1})\theta(E_{\rm min}-k_{2})
...\theta(E_{\rm min}-k_{n})\eqno(A4)$$
This calculation is discussed by Weinberg [7] and gives a probability
$$\left({E_{\rm min}\over \Lambda}\right)^{A}\ b(A)\eqno(A6)$$
where $b(A)\approx 1-(\pi A)^{2}/12+...$  is a different function that the
$1/\Gamma(1+A)$ that occurs in (A4).

\bigskip

\centerline{APPENDIX B: COMMENT ON A PREVIOUS CALCULATION}

In a previous paper on the semiclassical approximation [9], I computed the
residual effects of
infrared cancellation on processes  which occur in a finite-volume heat bath
but with no photons
detected outside the heat bath. The absence of  high energy photons  means none
were
radiated by the charge. The absence of low energy photons, which have a short
mean free path,
is unavoidable because  all those radiated by the charge would be thermalized
in the heat bath and
lose their identity. Consequently there is a threshold energy $\epsilon$ that
depends on the size of
the system. In the notation of Sec. II, the photons with $k<\epsilon$
contribute
$$\eqalign{R_{\epsilon}(z)=\int_{0}^{\epsilon}{d^{3}k\over
2k(2\pi)^{3}}&J_{\mu}(k)J^{\mu}(k)\cr
&\times\left([1+N]e^{ikz}+Ne^{-ikz}\right)\cr}\eqno(B1)$$
There is no constraint on the total energy $\omega$ so the probability of no
detected photons is
$$\int_{0}^{\infty}d\omega\ {dP_{\epsilon}\over
d\omega}=\exp[\overline{R}_{\epsilon}(0)]\eqno(B2)$$
where
$$\eqalign{\overline{R}_{\epsilon}(0)&=V+R_{\epsilon}(0)\cr
&=-A\int_{\epsilon}^{\infty}{dk\over k}[1+2N]\ e^{-k/\Lambda}\cr}\eqno(B3)$$
If $\epsilon\ll T$ then  $\overline{R}_{\epsilon}(0)\approx -2AT/\epsilon$.
However, for reasonable
mean free paths it invariably turns out that $\epsilon>T$, which made these
thermal corrections
small.

\vfill\eject
\centerline{REFERENCES}

\item{1.} D.R. Yennie, S.C. Frautschi, and H. Suura, Ann. of Phys. (NY), {\bf
13}, 379 (1961).

\item{2.} C. Itzykson and J.B. Zuber, {\sl Quantum Field Theory} (McGraw-Hill,
New York, 1980).

\item{3.} S. Weinberg, Phys. Rev. {\bf 135}, B1049 (1965).

\item {4.} A.J. Niemi and G.W. Semenoff, Ann. of Phys. (NY) {\bf 152}, 105
(1984).

\item{5.} N.P.Landsman and Ch. G. van Weert, Phys. Rep. {\bf 145}, 141 (1987).

\item{6.} R. J. Rivers,{\sl Path Integral Methods in Quantum Field Theory}
(Cambridge Univ. Press,
Cambridge Eng., 1987).

\item{7.} S.W. Weinberg, Phys. Rev.{140}, B516 (1965).

\item{8.} I.S. Gradshteyn and I.M. Ryzhik, {\sl Tables of Integrals, Series,
and Products}

\item{9.} H.A. Weldon, Phys. Rev. {D 44}, 3955 (1991); and
{\sl Hot Summer Daze: BNL Summer Study on QCD at Nonzero Temperature and
Density}, ed. A. Gocksch
and R. Pisarski (World Scientific, Singapore, 1992), p. 226.
Academic Press, New York, 1980).

\vskip1truein
\centerline{FIGURE CAPTIONS}

\bigskip

\item{Fig. 1} The dimensionless probability, $TdP/d\omega$, for a charged
particle
to radiate a net energy $\omega$. The first order result fails when $\omega<
A\pi T$.
For $\omega\ll A\pi T$ the exact result is constant, $TdP/d\omega\approx
1/A\pi^{2}$.
In this plot $A=0.01$, which corresponds to a momentum transfer $Q/m=3$.

\vskip0.5truein

\item{Fig. 2} Location of the branch cuts of the function
$\exp(\overline{R}(z))$
at $z=-i/\Lambda- in/T$. For $\omega >0$
the $z$
 integration contour may be wrapped around the branch branch cuts in the lower
half-plane.

\end

\end